\documentclass{sf2a-conf2014}
\usepackage{graphicx}
\usepackage{hyperref}
\usepackage[]{natbib}  
\usepackage{epstopdf}

\def\BibTeX{{\rm B\kern-.05em{\sc i\kern-.025em b}\kern-.08em
    T\kern-.1667em\lower.7ex\hbox{E}\kern-.125emX}}
\bibpunct{(}{)}{;}{a}{}{,}  

\begin{document}

\TitreGlobal{SF2A 2014}


\title{Ultra-weak magnetic fields in Am stars: $\beta$ UMa and $\theta$ Leo}

\runningtitle{Ultra-weak magnetic fields in Am stars}

\author{A. Blaz\`ere$^{1,}$}\address{LESIA, Observatoire de Paris, UMR 8109 du CNRS, UPMC, Universit\'e Paris-Diderot, 5 place Jules Janssen, 92195 Meudon, France}
\address{Universit\'e de Toulouse \& CNRS, Institut de Recherche en Astrophysique et Plan\'etologie, Toulouse, France}
\author{P. Petit$^{2}$}

\author{F. Ligni\`eres$^2$}

\author{M. Auri\`ere$^2$}

\author{T. B\"ohm$^2$}

\author{G. Wade}\address{Department of Physics, Royal Military College of Canada, PO Box 17000 Station Forces, Kingston, ON K7K 0C6, Canada}




\setcounter{page}{237}


\maketitle


\begin{abstract}
An extremely weak circularly-polarized signature was recently discovered in spectral lines of the chemically peculiar Am star Sirius A (Petit et al. 2011). This signal was interpreted as a Zeeman signature related to a sub-gauss longitudinal magnetic field, constituting the first detection of a surface magnetic field in an Am star. We present here ultra-deep spectropolarimetric observations of two other bright Am stars, $\beta$ UMa and $\theta$ Leo, observed with the NARVAL spectropolarimeter. The line profiles of the two stars display circularly-polarized signatures similar in shape to the observations gathered for Sirius A. These new detections suggest that very weak magnetic fields may be present in the photospheres of a significant fraction of intermediate-mass stars, although the strongly asymmetric Zeeman signatures measured so far in Am stars (featuring a prominent positive lobe and no detected negative lobe) are not expected in the standard theory of the Zeeman effect.
\end{abstract}



\section{Introduction}
Magnetic fields play an important role in the evolution of hot stars (O, B and A stars). However, the origin and even the basic properties of hot star magnetic fields are still poorly understood. About 7\% of hot stars are found to be strongly magnetic with a longitudinal magnetic field in excess of 100 G \citep{wade13}. 
But recently, a sub-gauss longitudinal magnetic field has been discovered in the normal A star Vega \citep{lignieres09}. This detection raises the question of the ubiquity of magnetic fields in objects belonging to this mass domain. 
In 2011, another polarimetric signal was detected in the bright Am star Sirius A \citep{petit11}. For this object, the polarized signature in circular polarization is not of null integral over the line profile as in other massive stars, since the Stokes V line profile exhibits a positive lobe dominating over the negative one (in amplitude and area). 
Here, we present the results of a magnetic field search carried out for two other bright Am stars: $\beta$ UMa and $\theta$ Leo. The fundamental parameters of both targets are presented in Table~\ref{parameter}.
The Am stars are  chemically peculiar stars exhibiting overabundances of iron-group elements such as zinc, strontium, zirconium and barium and deficiencies of others such as calcium and scandium. Most Am stars also feature low projected rotational velocities, as compared to normal A stars \citep{abt09}.

\begin{table}[h]
\label{parameter}
\caption{Fundamental parameters of $\beta$ UMa and $\theta$ Leo}
\centering
\begin{tabular}{c c c}
\hline
  & $\beta$ UMa & $\theta$ Leo\\
\hline
\hline
   spectral type & A1V & A2V \\
   $T_{eff}$    & 9600 K$^{a}$ & 9350 K$^{b}$  \\
   log g   &   3.83$^{c}$   & 3.65$^{b}$  \\
   Mass & 2.7 $M_{\odot}^{d}$ & 2.5 $M_{\odot}^{d}$\\
   Radius & 3 $R_{\odot}^{a}$ & 4.3 $R_{\odot}^{a}$\\
   vsini & 23 km/s$^{e}$ & 46 km/s$^{e}$\\
\hline
\multicolumn{2}{l}{$^{a}$ \cite{boyajian12}} & $^{c}$ \cite{monier05}\\
\multicolumn{2}{l}{$^{b}$ \cite{smith93}}&$^{d}$\cite{wyatt07}\\
$^{e}$ \cite{royer02}\\
\end{tabular}
\end{table}

\section{Data analysis}
Data were taken with the NARVAL spectropolarimeter. Narval
is operated at the 2-meter Bernard Lyot Telescope (TBL), at the
summit of Pic du Midi in the French Pyr\'en\'ees. This fibre-fed
spectropolarimeter was especially designed and optimized to detect stellar surface magnetic
fields through the polarization they generate in photospheric lines and provides complete coverage of the
optical spectrum from 3700 to 10500 $\AA$ on 40 echelle orders with a spectral
resolution of 65000.

$\beta$ UMa was observed in March/April 2010 and March/April 2011, while observations of $\theta$ Leo were collected in  January/March/April 2012, March/April 2013 and May 2014 (see Table~\ref{obs}).

\begin{table}[h]
\label{obs}
\caption{Journal of observations}
\centering
\begin{tabular}{c c c c}
\hline
date & mid-HJD & star & $T_{exp}$ (s)\\
\hline
\hline
17mar10 & 2455273.52016 & $\beta$ UMa & 16 $\times$ 4 $\times$ 107\\
06apr10 & 2455293.41173 & $\beta$ UMa & 17 $\times$ 4 $\times$ 107\\
10apr10 & 2455297.44406 & $\beta$ UMa & 19 $\times$ 4 $\times$ 107\\
11apr10 & 2455298.39660 & $\beta$ UMa & 19 $\times$ 4 $\times$ 107\\
25mar11 & 2455646.42619 & $\beta$ UMa & 25 $\times$ 4 $\times$ 107\\
31mar11 & 2455652.50367 & $\beta$ UMa & 25 $\times$ 4 $\times$ 107\\
02apr11 & 2455654.37913 & $\beta$ UMa & 03 $\times$ 4 $\times$ 107\\
04apr11 & 2455656.46150 & $\beta$ UMa & 24 $\times$ 4 $\times$ 107\\
22jan12 & 2455949.64440 & $\theta$ Leo & 05 $\times$ 4 $\times$ 180\\
23jan12 & 2455950.62844 & $\theta$ Leo & 05 $\times$ 4 $\times$ 180\\
24jan12 & 2455951.62417 & $\theta$ Leo & 05 $\times$ 4 $\times$ 180\\
25jan12 & 2455952.64032 & $\theta$ Leo & 05 $\times$ 4 $\times$ 180\\
14mar12 & 2456001.57862 & $\theta$ Leo & 05 $\times$ 4 $\times$ 180\\
15mar12 & 2456002.52449 & $\theta$ Leo & 10 $\times$ 4 $\times$ 180\\
24mar12 & 2456011.52572 & $\theta$ Leo & 05 $\times$ 4 $\times$ 180\\
25mar12 & 2456012.50225 & $\theta$ Leo & 05 $\times$ 4 $\times$ 180\\
27mar12 & 2456013.39956 & $\theta$ Leo & 10 $\times$ 4 $\times$ 180\\
21mar13 & 2456373.48791 & $\theta$ Leo & 09 $\times$ 4 $\times$ 180\\
23mar13 & 2456375.46496 & $\theta$ Leo & 09 $\times$ 4 $\times$ 180\\
16apr13 & 2456399.44433 & $\theta$ Leo & 09 $\times$ 4 $\times$ 180\\
17apr13 & 2456400.49237 & $\theta$ Leo & 09 $\times$ 4 $\times$ 180\\
22apr13 & 2456405.51179 & $\theta$ Leo & 09 $\times$ 4 $\times$ 180\\
23apr13 & 2456406.45400 & $\theta$ Leo & 09 $\times$ 4 $\times$ 180\\
24apr13 & 2456407.50188 & $\theta$ Leo & 09 $\times$ 4 $\times$ 180\\
14apr14 & 2456762.44478 & $\theta$ Leo & 05 $\times$ 4 $\times$ 180\\
07may14 & 2456785.40762 & $\theta$ Leo & 05 $\times$ 4 $\times$ 180\\
08may14 & 2456786.41135 & $\theta$ Leo & 05 $\times$ 4 $\times$ 180\\
09may14 & 2456787.41580 & $\theta$ Leo & 05 $\times$ 4 $\times$ 180\\
14may14 & 2456792.47117 & $\theta$ Leo & 05 $\times$ 4 $\times$ 180\\
15may14 & 2456793.41300 & $\theta$ Leo & 05 $\times$ 4 $\times$ 180\\

\end{tabular}
\end{table}

To test whether $\beta$ UMa and $\theta$ Leo are magnetic, we applied the well-known and commonly used Least-Squares Deconvolution (LSD) technique \citep{donati97} on each spectrum of both stars and computed LSD pseudo line profiles from all available photospheric lines. The line lists used are created from a list of lines extracted from the VALD data base \citep{piskunov95, kupka99} using the respective effective temperature and log g of both stars (Table~\ref{parameter}). To further improve the signal-to-noise ratio, we then coadded all LSD profiles of each star, resulting in one single averaged LSD profile for each target. The result are shown in Figure~\ref{lsd}.

\section{Results}

\begin{figure}[!t]
\begin{center}
\includegraphics[scale=0.28]{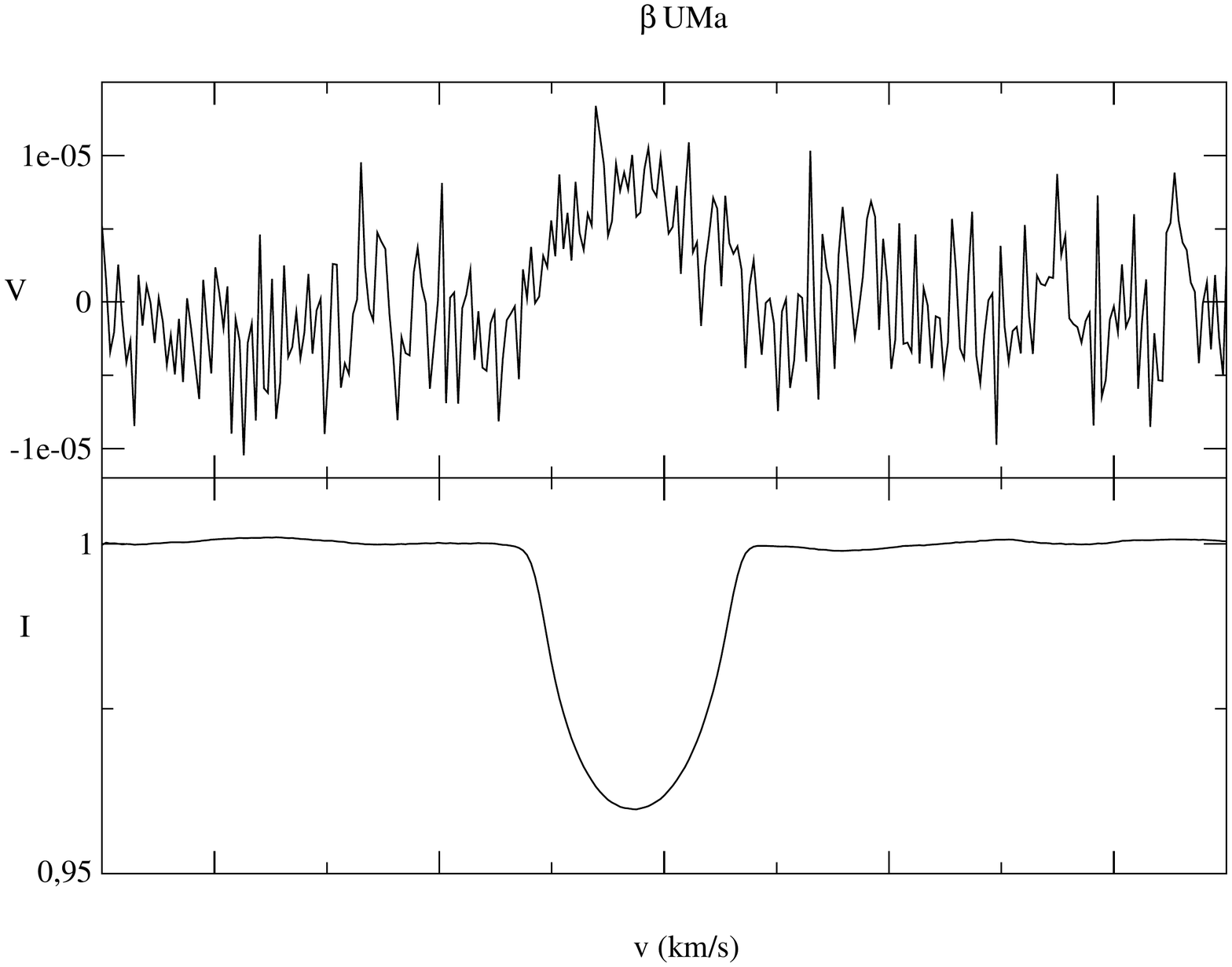}
\includegraphics[scale=0.28]{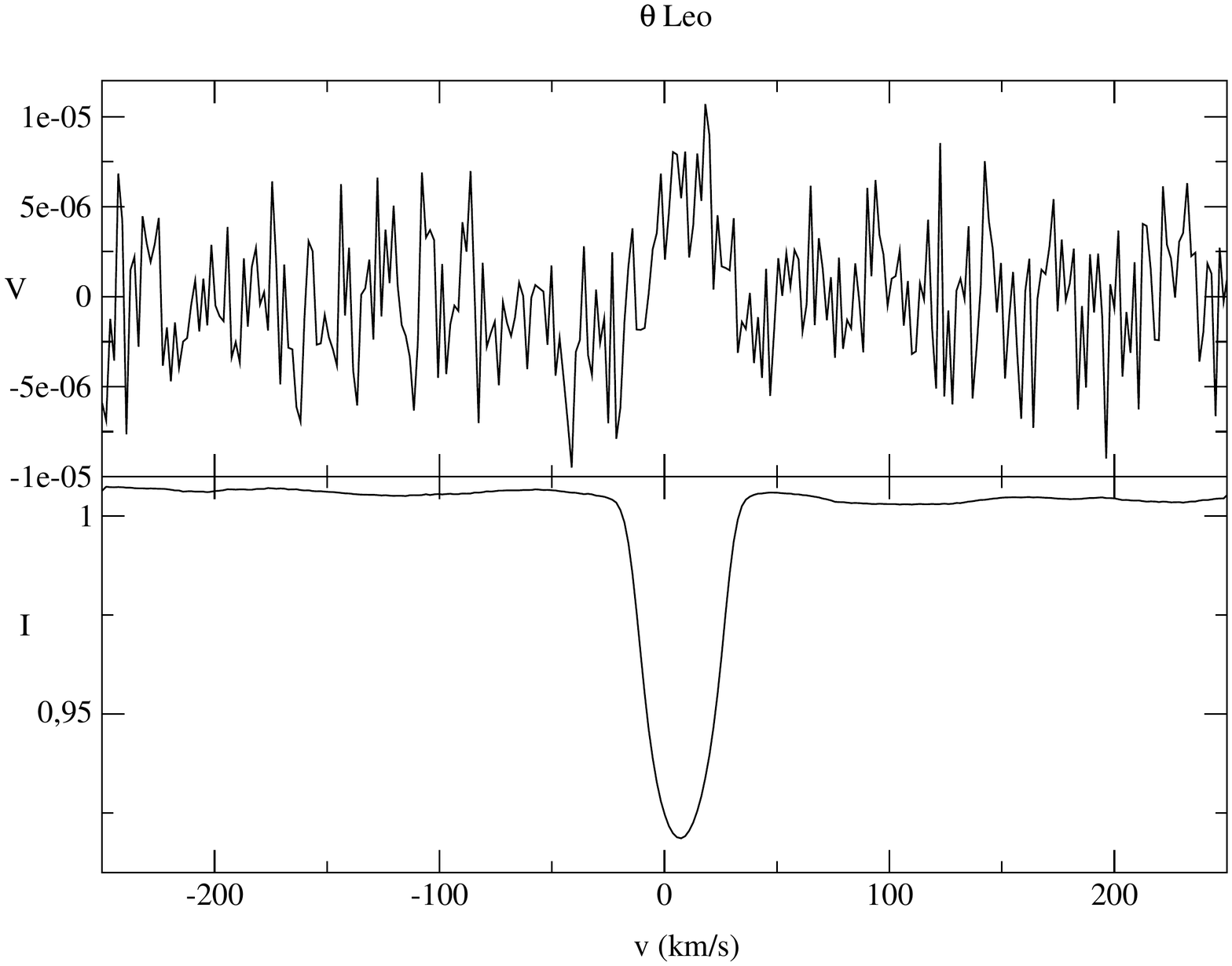}
\caption{Left: normalized Stokes I and V averaged LSD profiles for $\beta$ UMa. Right: same figure for $\theta$ Leo.}
\label{lsd}
\end{center}
\end{figure}

The line profiles of the two stars display circularly-polarized signatures similar in shape (see Figure~\ref{lsd}) to the observations previously gathered for Sirius A. We have also separately coadded LSD profiles for each observing year (not shown here) to evaluate the temporal stability of this signal, concluding that the signatures are stable over the time-span of our observations. 

\section{Conclusions}
These new detections may be interpreted to suggest that sub-gauss magnetic fields are present in the photosphere of a significant
fraction of intermediate-mass stars, although the strongly asymmetric Zeeman signatures measured so far in Am stars (featuring a prominent positive lobe and essentially no negative lobe) are not expected in the standard theory of the Zeeman effect (see possible interpretations in \citealt{petit11}). New observations are currently being carried out to gain a better statistics of the prevalence  of weak magnetic fields in intermediate-mass stars and evaluate the impact of various stellar parameters on the Vega-like magnetism.


\begin{acknowledgements}
We acknowledge support from the ANR (Agence Nationale de la Recherche)
project Imagine. This research has made use of the SIMBAD database operated
at CDS, Strasbourg (France), and of NASA's Astrophysics Data System (ADS).

\end{acknowledgements}

\bibliographystyle{aa}  
\bibliography{biblio} 

\end{document}